\documentclass[aps,prc,twocolumn,floatfix,amsmath,fleqn,superscriptaddress]{revtex4-1}
\usepackage[utf8]{inputenc}
\usepackage[T1]{fontenc}

\usepackage{xcolor}
\usepackage{graphicx}

\usepackage{scalefnt}

\usepackage{amsmath}
\usepackage{amssymb}
\usepackage{times,txfonts}
\usepackage{pifont}

\newcommand{\la}{\langle}
\newcommand{\ra}{\rangle}

\newcommand{\symboldiamond}[1][black]{{\color{#1}$\blacklozenge$}}
\newcommand{\symboltriangle}[1][black]{{\color{#1}\ding{115}}}

\newcommand{\symbolbox}[1][black]{{\color{#1}\scalefont{0.75}$\blacksquare$}}
\newcommand{\symbolcircle}[1][black]{{\color{#1}\scalefont{0.75}\ding{108}}}

\newcommand{\symbolcross}[1][black]{{\color{#1}\ding{58}}}

\definecolor{FGViolet}{rgb}{0.61,0.32,0.61}
\definecolor{FGDarkBlue}{rgb}{0,0,0.6}
\definecolor{FGBlue}{rgb}{0,0,0.8}
\definecolor{FGLightBlue}{rgb}{0.2, 0.6, 0.8}
\definecolor{FGGreen}{rgb}{0.2,0.7,0.2}
\definecolor{FGLightGreen}{rgb}{0.4,1,0.4}
\definecolor{FGYellow}{rgb}{1,0.95,0}
\definecolor{FGOrange}{rgb}{0.95,0.5,0.1}
\definecolor{FGRed}{rgb}{0.8,0,0}
\definecolor{FGWhite}{rgb}{1,1,1}
\definecolor{FGLightGray}{rgb}{0.8,0.8,0.8}
\definecolor{FGGray}{rgb}{0.5,0.5,0.5}
\definecolor{FGDarkGray}{rgb}{0.3,0.3,0.3}
\definecolor{FGBlack}{rgb}{0,0,0}

\newcommand{\elem}[2]{\ensuremath{{}^{#2}\text{#1}}}

\newcommand{\ame}[2]{\ensuremath{  H_{#1,#2} }}

\newcommand{\pr}{\prime}

\begin{document}

\title{Natural orbitals for \emph{ab initio} no-core shell model calculations}

\author{Alexander Tichai}
\email{alexander.tichai@cea.fr}
\affiliation{Institut f\"ur Kernphysik, Technische Universit\"at Darmstadt, Schlossgartenstr.\ 2, 64289 Darmstadt, Germany}
\affiliation{ESNT, IRFU, CEA, Universit\'e Paris-Saclay, F-91191 Gif-sur-Yvette, France}

\author{Julius M\"uller}
\email{julius.mueller@physik.tu-darmstadt.de}

\author{Klaus Vobig}
\email{klaus.vobig@physik.tu-darmstadt.de}

\author{Robert Roth}
\email{robert.roth@physik.tu-darmstadt.de}
\affiliation{Institut f\"ur Kernphysik, Technische Universit\"at Darmstadt, Schlossgartenstr.\ 2, 64289 Darmstadt, Germany}

\date{\today}

\begin{abstract}
We explore the impact of optimizations of the single-particle basis on the convergence behavior and robustness of \emph{ab initio} no-core shell model calculations of nuclei. Our focus is on novel basis sets defined by the natural orbitals of a correlated one-body density matrix that is obtained in second-order many-body perturbation theory. Using a perturbatively improved density matrix as starting point informs the single-particle basis about the dominant correlation effects on the global structure of the many-body state, while keeping the computational cost for the basis optimization at a minimum. Already the comparison of the radial single-particle wavefunctions reveals the superiority of the natural-orbital basis compared to a Hartree-Fock or harmonic oscillator basis, and it highlights pathologies of the Hartree-Fock basis. We compare the model-space convergence of energies, root-mean-square radii, and selected electromagnetic observables for all three basis sets for selected p-shell nuclei using chiral interactions with explicit three-nucleon terms. In all cases the natural-orbital basis provides the fastest and most robust convergence, making it the most efficient basis for no-core shell model calculations. As an application we present no-core shell model calculations for the ground-state energies of all oxygen isotopes and assess the accuracy of the normal-ordered two-body approximation of the three-nucleon interaction in the natural-orbital basis.   

\end{abstract}


\maketitle

\section{Introduction}

The \emph{ab initio} solution of the quantum many-body problem using realistic nuclear interactions poses a long-standing challenge in modern nuclear structure theory. 
Configuration interaction (CI) approaches provide a simple tool to access a variety of nuclear observables via the solution of a large-scale matrix eigenvalue problem for the Hamiltonian represented in a many-body basis.
Of particular importance are the no-core CI approaches, with the \emph{ab initio} no-core shell-model (NCSM) as a prime representative, which treat all nucleons as active degrees of freedom \cite{BaVa13,NaQu09}.
In its original formulation, the NCSM uses the harmonic oscillator (HO) single-particle basis in conjunction with a truncation of the many-body Slater-determinant basis in terms of the total number of HO excitation quanta, $N_\text{max}$.
Due to its conceptual simplicity, the NCSM can be straightforwardly applied to the calculation of a full suite of observables for ground and low-lying excited states of all light nuclei.
However, it is well-known that the HO single-particle wavefunctions do not posses the correct asymptotic behavior and that convergence of energies and other observables with increasing $N_\text{max}$ can be slow and strongly dependent on the oscillator frequency $\hbar \Omega$. Convergence is particularly difficult for observables sensitive to the long-range behavior of the many-body wavefunction, such as the root-mean-square radius and the electric quadrupole moment or transition strength. 

The question arises, to what extend this convergence can be accelerated by an optimization of the single-particle basis and how such an optimization can be performed. An obvious choice for a basis optimization method is the Hartree-Fock (HF) approximation, i.e., a variational calculation for the ground state of a given nucleus using a single Slater determinant as a trial state, whose single-particle states serve as variational degrees of freedom \cite{RiSc80,RoPa06}. These HF-optimized single-particle basis sets are widely used in methods targeting medium-mass nuclei, such as in-medium similarity renormalization group \cite{H15,He17}, coupled-cluster \cite{HaHj16,HaPa14}, or self-consistent Green's function methods \cite{BaCa17,CiBa13}. An obvious downside of this optimization strategy is that the single-particle basis targets the completely uncorrelated mean-field approximation for the nucleus of choice. Depending on the interaction, the HF ground-state might be not even bound, so the HF single-particle basis accounts for a completely different, potentially unphysical situation. Even if soft interactions are being used, the HF ground state lacks correlation effects and the binding energy is underestimated significantly. Therefore, the HF-optimized single-particle basis might not be the best starting point for the description of the fully correlated nucleus.

A more general strategy for constructing an optimized single-particle basis are the so-called natural orbitals, i.e., the eigenvectors of the one-body density matrix. This one-body density matrix is obtained from a separate many-body calculation, which might include correlations at different levels of approximation. It is known from applications in quantum chemistry and atomic physics that natural orbitals provide an excellent single-particle basis to speed up the convergence of the CI expansion~\cite{BeDav66}.

If using a HF calculation to construct the one-body density matrix of the ground state of the target nucleus, thus ignoring correlations, then the natural orbitals are identical to the HF single-particle basis. The other extreme would be a fully correlated NCSM calculation in a large model space to construct the one-body density matrix, yielding a natural orbital basis adapted to the global properties of the fully correlated many-body state. This strategy was pioneered for light nuclear systems in the NCSM context in Ref.~\cite{CoCa16}. An obvious drawback is the extra computational effort for the NCSM calculation underlying the natural orbitals.

Prior investigations in quantum chemistry have shown that approximate natural orbitals perform equally well in CI applications, but require much lower computational effort for their construction~\cite{Ha73,SiHa74}. Therefore, we develop a simplified strategy to determine an optimized natural-orbital basis for NCSM-type calculations. We use a second-order corrected one-body density matrix that is conveniently derived using Hartree-Fock many-body perturbation theory (MBPT)~\cite{ShBa09,SzOs82,Ti16,RoPa06}. Subsequently, we compare HO, HF and natural orbital radial wavefunctions and explore the impact of the different bases on the model-space convergence of nuclear observables in large-scale NCSM calculations.

\section{Natural Orbitals from MBPT}

We aim at the construction of the one-body density matrix
\begin{align}
\rho_{pq} = \la \Psi \vert\, c_p^\dagger c_q\, \vert \Psi \ra\, ,
\end{align}
for a nuclear many-body state $\vert \Psi \ra$, the so-called reference state, that includes correlations at least in an approximate way. 
Here and in the following $c_p^\dagger, c_p$ denote single-particle creation and annihilation operators, respectively, for an auxiliary single-particle basis. The index $p$ refers to the set of quantum numbers $(n_p, l_p, j_p, m_p, t_p)$, where $n_p$ is the radial quantum number, $l_p$ the orbital angular momentum, $j_p$ the total angular momentum with projection $m_p$, and $t_p$ the isospin projection. 

Instead of performing large-scale NCSM or CI calculations to obtain $\vert \Psi \ra$, we use a computationally more efficient alternative based on MBPT. We approximate the eigenstate up to second order in the MBPT expansion 
\begin{align}
\vert \Psi \ra \approx \vert \Psi^{(0)} \ra + \vert \Psi^{(1)} \ra + \vert \Psi ^{(2)} \ra\, , 
\end{align}
where $\vert \Psi^{(n)} \ra $ denotes the $n$-th order state correction on top of the unperturbed state $|\Psi^{(0)}\ra$. Building on our previous work on HF-MBPT, its order-by-order convergence, and the quality of low-order approximations, we use an angular-momentum restricted HF ground state as unperturbed state $|\Psi^{(0)}\ra = | \Phi_\text{HF}\ra$ and formulate the  M{\o}ller-Plesset partitioning of the intrinsic nuclear Hamiltonian $H$ accordingly. For more details we refer the reader to Ref. \cite{Ti16}. 

Following Refs.~\cite{StBa73,RoPa06}, we can write the one-body density matrix up to second order in the interaction as 
\begin{align}
\label{eq:diags}
\rho \approx \rho^{(00)} + \rho^{(02)} + \rho^{(20)} + \rho^{(11)}\, , 
\end{align}
where $\rho^{(00)}$ denotes the zeroth-order HF density matrix and
\begin{align}
\rho^{(02)}_{pq} &= \la \Psi^{(0)} \vert\,  c_p^\dagger c_q\, \vert \Psi^{(2)} \ra = \rho^{(20)}_{qp} , \\
\rho^{(11)}_{pq} &= \la \Psi^{(1)} \vert\, c_p^\dagger c_q\, \vert \Psi^{(1)} \ra \, . 
\end{align}
These nontrivial corrections to the density matrix can be conveniently written as
\begin{align}
\rho^{(02)} &= D^\text{(A)} + D^\text{(B)}, \\
\rho^{(11)} &= D^\text{(C)} + D^\text{(D)},
\end{align}
where the individual terms are given by
\begin{align}
D^\text{(A)}_{i^\pr a^\pr} &= \phantom{-}\frac{1}{2}\sum_{abi}  \frac{\ame{i^\pr i}{ab} \ame{ab}{a^\pr i} } { (\epsilon_{i^\pr} - \epsilon_{a^\pr}) (\epsilon_{i^\pr} + \epsilon_{i} - \epsilon_{a} - \epsilon_{b} )}, \\
D^\text{(B)}_{i^\pr a^\pr} &= -\frac{1}{2}\sum_{aij} \frac{\ame{i^\pr a}{ij} \ame{ij}{a^\pr a} } { (\epsilon_{i^\pr} - \epsilon_{a^\pr}) (\epsilon_{i} + \epsilon_{j} - \epsilon_{a^\pr} - \epsilon_{a} )}, \\
D^\text{(C)}_{i^\pr j^\pr} &= - \frac{1}{2} \sum_{abi} \frac{\ame{i^\pr i}{ab} \ame{ab}{j^\pr i} } { (\epsilon_{i^\pr} + \epsilon_{i} - \epsilon_a - \epsilon_b) (\epsilon_{j^\pr} + \epsilon_{i} - \epsilon_{a} - \epsilon_{b} )}, \\
D^\text{(D)}_{a^\pr b^\pr } &= \phantom{-} \frac{1}{2} \sum_{aij} \frac{\ame{a^\pr a}{ij} \ame{ij}{b^\pr a} } { (\epsilon_{i} + \epsilon_{j} - \epsilon_{a^\pr} - \epsilon_{a}) (\epsilon_{i} + \epsilon_{j} - \epsilon_{b^\pr} - \epsilon_{a} )}.
\end{align}
These expressions make use of the particle-hole formalism: Single-particle states occupied in the unperturbed HF state are denoted by indices $i,j,...$ (holes), unoccupied states by indices $a,b,...$ (particles). Furthermore, $\ame{pq}{rs}$ denote antisymmetric two-body matrix elements of the intrinsic Hamiltonian and $\epsilon_p$ the HF single-particle energies. 

For reasons of computational efficiency, we choose the simplest possible implementation of MBPT assuming a two-body Hamiltonian and a single-determinant reference state. For the inclusion of three-body interactions we apply the normal-ordered two-body approximation with respect to the single-determinantal HF reference state \cite{RoBi12,BiLa13}. Note that the HF calculation itself uses the HO as computational basis and all initial matrix elements are specified in HO representation and subsequently transformed to the HF basis. For open-$j$-shell systems we use a symmetry-constrained HF determinant constructed in an equal-filling approximation, thus, yielding a HF solution with fractional occupation numbers in the degenerate shell. This also facilitates the derivation of an angular-momentum coupled form of the above equations, which is used in all following calculations.

We note that there is no contribution $ \la \Psi^{(0)} \vert\,  c_p^\dagger c_q \,\vert \Psi^{(1)} \ra$ since Brillouin's theorem prevents single-excitations in the first-order state correction from direct mixing with the HF ground state.
Furthermore, the trace of the correlated second-order density matrix is the same as the trace of the HF density, since
\begin{align}
\sum_i D_{ii}^\text{(C)} + \sum_a D_{aa}^\text{(D)} =0\, .
\end{align}

Due to the symmetries of the Hamiltonian, the resulting correlated density matrix is block-diagonal in $l$, $j$, $t$ and $m$ and it is independent of $m$ due to our symmetry constraint. Therefore, we obtain a set of spherical natural orbitals by diagonalizing the one-body density matrix in each $(ljt)$ block. The new natural orbital single-particle states (NAT) are then given by superposition of the HF or HO basis states according to
\begin{align}
\vert nljmt \ra_{\text{NAT}} 
&= \sum_{n^\pr} c^{(ljt)}_{n n^\pr} \;\vert n^\pr ljmt\ra_{\text{HF}} \\
&= \sum_{n^\pr} \tilde{c}^{(ljt)}_{n n^\pr} \;\vert n^\pr ljmt\ra_{\text{HO}}\, ,
\label{eq:trafo}
\end{align}
where $c^{(ljt)}_{n n^\pr}$ are the expansion coefficients with respect to the HF basis obtained from the diagonalization of the one-body density matrix, and $\tilde{c}^{(ljt)}_{n n^\pr}$ are the expansion coefficients with respect to the HO basis that encapsulate the basis transformation between HO and HF in addition. Since we use a truncated HO set as computational basis for the construction of the HF and NAT basis sets, these basis sets also contain the oscillator frequency $\hbar\Omega$ as a parameter.

The expansion \eqref{eq:trafo} is used to transform the matrix elements of all relevant operators, e.g., nucleon-nucleon (NN) and three-nucleon (3N) interactions and a collection of electromagnetic operators, from the initial HO representation to the NAT basis. Those matrix elements are stored to disk for the subsequent many-body calculation. In the following we focus on applications in the NCSM, however, the same NAT matrix elements can be used, e.g., in in-medium similarity renormalization group or coupled-cluster calculations for medium-mass nuclei.  
Note that the computationally most demanding step in the whole procedure is actually the transformation of the various matrix elements to the NAT basis, particularly if explicit three-nucleon forces are included---as we do in the following. 

\begin{figure}[t]
\includegraphics[width=1.0\columnwidth]{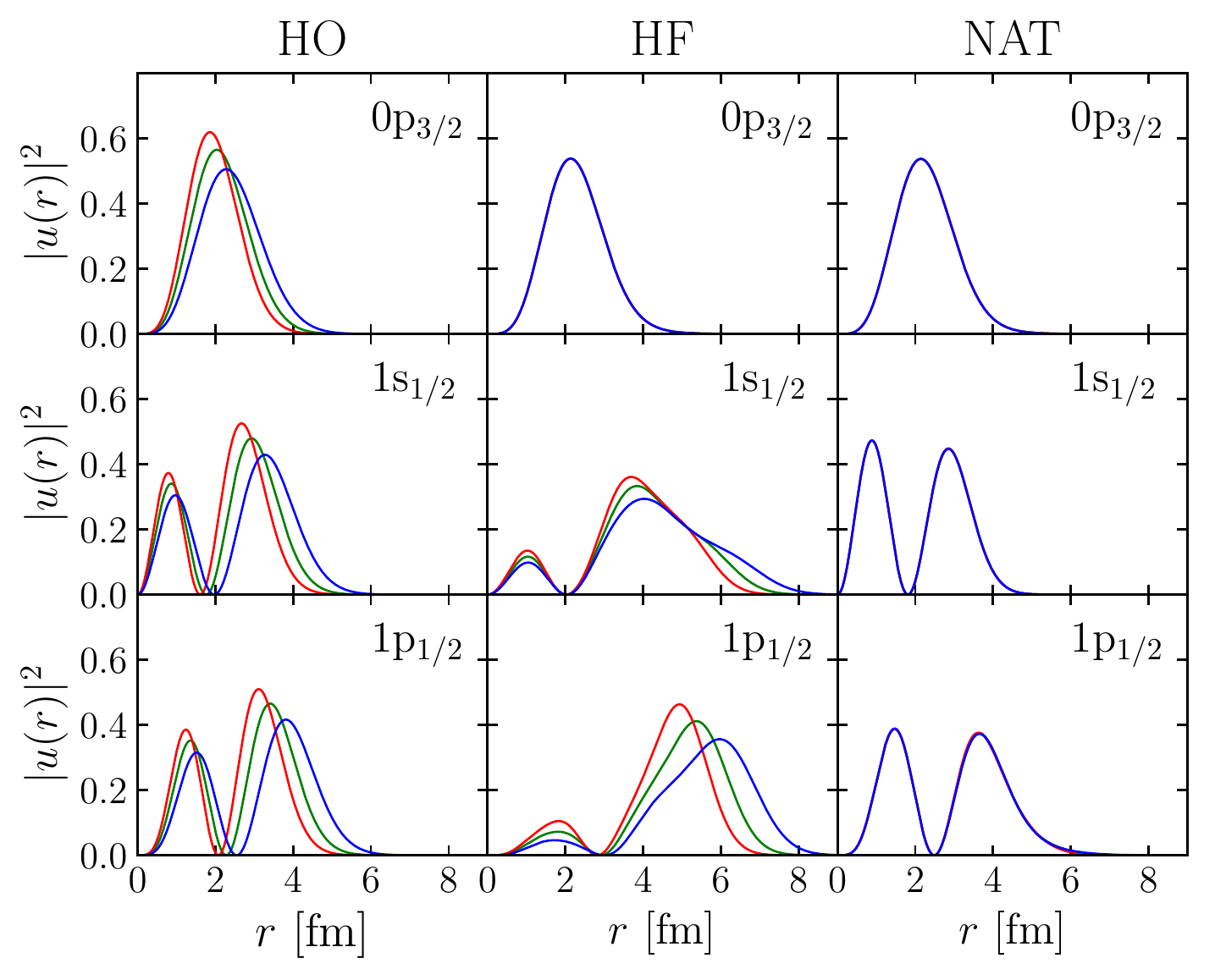}
\caption{(color online) Plot of the squared radial wavefunctions $\vert u(r)\vert^2$ (in units of $\text{fm}^{-1}$) of the HO basis (left-hand column), the HF basis (center column), and NAT basis (right-hand column) for \elem{O}{16}. The different rows display different single-particle orbits: the occupied $0p_{3/2}$ orbit (top row) and the unoccupied and HF-unbound $1s_{1/2}$ and $1p_{1/2}$ orbit (middle and bottom row). The different lines correspond to different values for the oscillator frequency $\hbar \Omega=16\,\text{MeV}$ (blue), $20\,\text{MeV}$ (green), and  $24\,\text{MeV}$ (red). We use the NN+3N interaction with $\Lambda_{\text{3N}}=400\,\text{MeV}/c$ and $\alpha = 0.08\, \text{fm}^4$.}
\label{fig:spwf}
\end{figure}

\section{Benchmark Calculations}

We explore the convergence properties of the NAT basis in comparison to conventional HF and HO basis sets for different observables in the framework of the NCSM. For all basis sets, we adopt the conventional $N_{\max}$ truncation, defined with respect to the radial and orbital angular momentum quantum numbers. For the HF and NAT basis this does not necessarily correspond to a truncation in the unperturbed excitation energy of the basis states anymore, but is still provides a very efficient truncation of the many-body basis that guarantees good angular momenta. With the HF and NAT bases we also sacrifice the exact factorisation of intrinsic and center-of-mass components of the eigenstates, which is formally guaranteed for the HO basis with an $N_{\max}$ truncation.   

For the following calculations we use nuclear Hamiltonians with NN and 3N interactions derived from chiral effective field theory. To facilitate the comparison with previous calculations and other many-body approaches, we use the classic NN interaction at next-to-next-to-next-to leading order by Entem and Machleidt~\cite{EnMa03} in combination with a local 3N force at next-to-next-to leading order with cutoffs of $\Lambda_{3\text{N}}=400$ or $500\,\text{MeV}/c$ depending on the system under consideration~\cite{Na07,RoBi12}. The Hamiltonian is transformed at the three-body level with the similarity renormalization group (SRG) approach \cite{BoFu07,HeRo07,RoRe08,RoLa11,JuMa13,RoCa13}, i.e., induced many-body forces beyond three-body level are discarded. We use a flow parameter $\alpha=0.08 \,\text{fm}^4$ as in previous studies \cite{RoLa11,RoBi12,HeBi13,BiLa14}.
The single-particle basis for the HF and NAT calculations contains all single-particle states up to $e_\text{max} = (2n +l)_\text{max} = 12$. For the matrix elements of the 3N interaction we use an additional truncation $e_1 + e_2 + e_3 \leq E_{3\text{max}} = 14$, which was shown to be sufficient for nuclei up into the sd-shell \cite{BiLa14}. For the oxygen isotopes beyond $N_{\max}=4$ we employ an importance truncation of the NCSM model space, as discussed in \cite{Roth09,RoLa11,MaVa14}.

\subsection{Single-Particle Wavefunctions} 

We start with a comparison of the structure of the radial wavefunctions in the HO, HF, and NAT single-particle bases in Fig.~\ref{fig:spwf}. We focus on the description of the closed-shell nucleus \elem{O}{16} and distinguish hole and particle states, i.e., occupied and unoccupied single-particle states in the HF determinant, respectively. The first row of Fig.~\ref{fig:spwf} depicts the $0p_{3/2}$ radial wavefunctions, corresponding to occupied states, for three different frequencies of the underlying HO basis. Trivially, all the HO wavefunctions show a simple radial scaling with the oscillator lengths proportional to $(\hbar\Omega)^{-1/2}$. The HF and NAT wavefunctions, however, are independent of the oscillator frequency and almost identical. The lower two rows of Fig.~\ref{fig:spwf} show the $1s_{1/2}$ and $1p_{1/2}$ wavefunctions as representatives of unoccupied and HF-unbound single-particle states. While the NAT wavefunctions are again completely frequency independent and similar in shape to the HO wavefunctions, the HF wavefunctions exhibits severe pathologies---they show unphysical distortions and a very pronounced frequency dependence, particularly in the long-range part.

The deficiencies of the HF wavefunctions for unoccupied states are not surprising, given that the self-consistent solution of the HF equations only provides a variational optimization of the occupied states, while unoccupied single-particle states are only fixed via orthogonality and normalization. This is different for the NAT basis, since all single-particle states---also the ones not occupied in HF---contribute to the correlated ground state and, thus, carry relevant physical information on the global structure of the ground state. As long as the Hamiltonian mixes the high-lying single-particle states into the ground state, the one-body density matrix and the natural orbitals provide an efficient tool for optimizing their wavefunctions.

\subsection{Ground-State Energies}

\begin{figure}[t!]
\centering
\includegraphics[width=1\columnwidth]{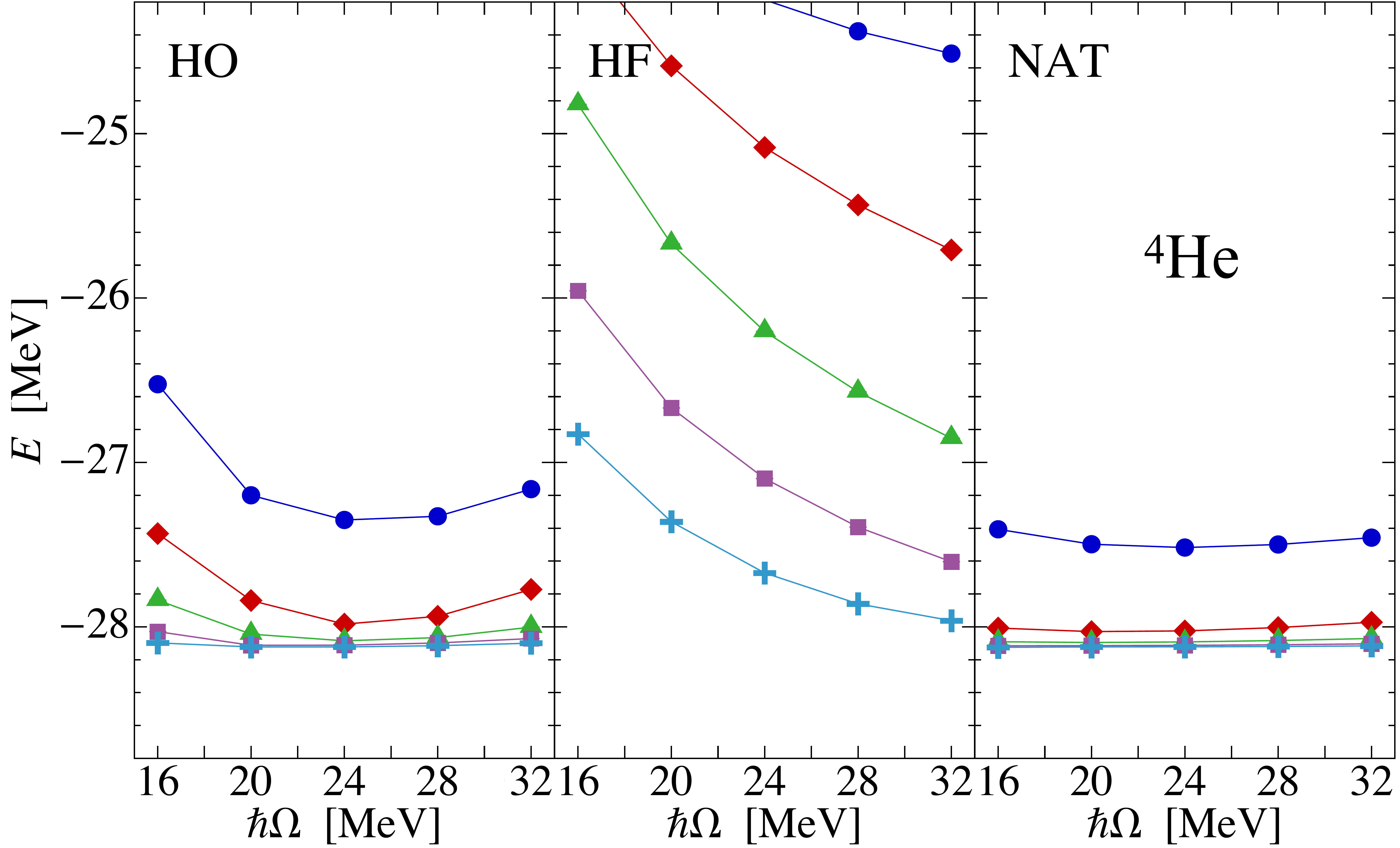}
\caption{(color online) Ground-state energies for \elem{He}{4} obtained in the NCSM with the HO, HF, and NAT basis sets (panels from left to right) as function of the oscillator frequency $\hbar\Omega$ for $N_{\max}=4$ (\symbolcircle[FGBlue]), $6$ (\symboldiamond[FGRed]), $8$ (\symboltriangle[FGGreen]), $10$ (\symbolbox[FGViolet]), and $12$ (\symbolcross[FGLightBlue]). All calculations employ the chiral NN+3N interaction ($\Lambda_{\text{3N}}=400\,\text{MeV}/c$) after an SRG evolution with $\alpha = 0.08\,\text{fm}^4$.}
\label{fig:He4egs}
\end{figure}

\begin{figure}[t!]
\centering
\includegraphics[width=1\columnwidth]{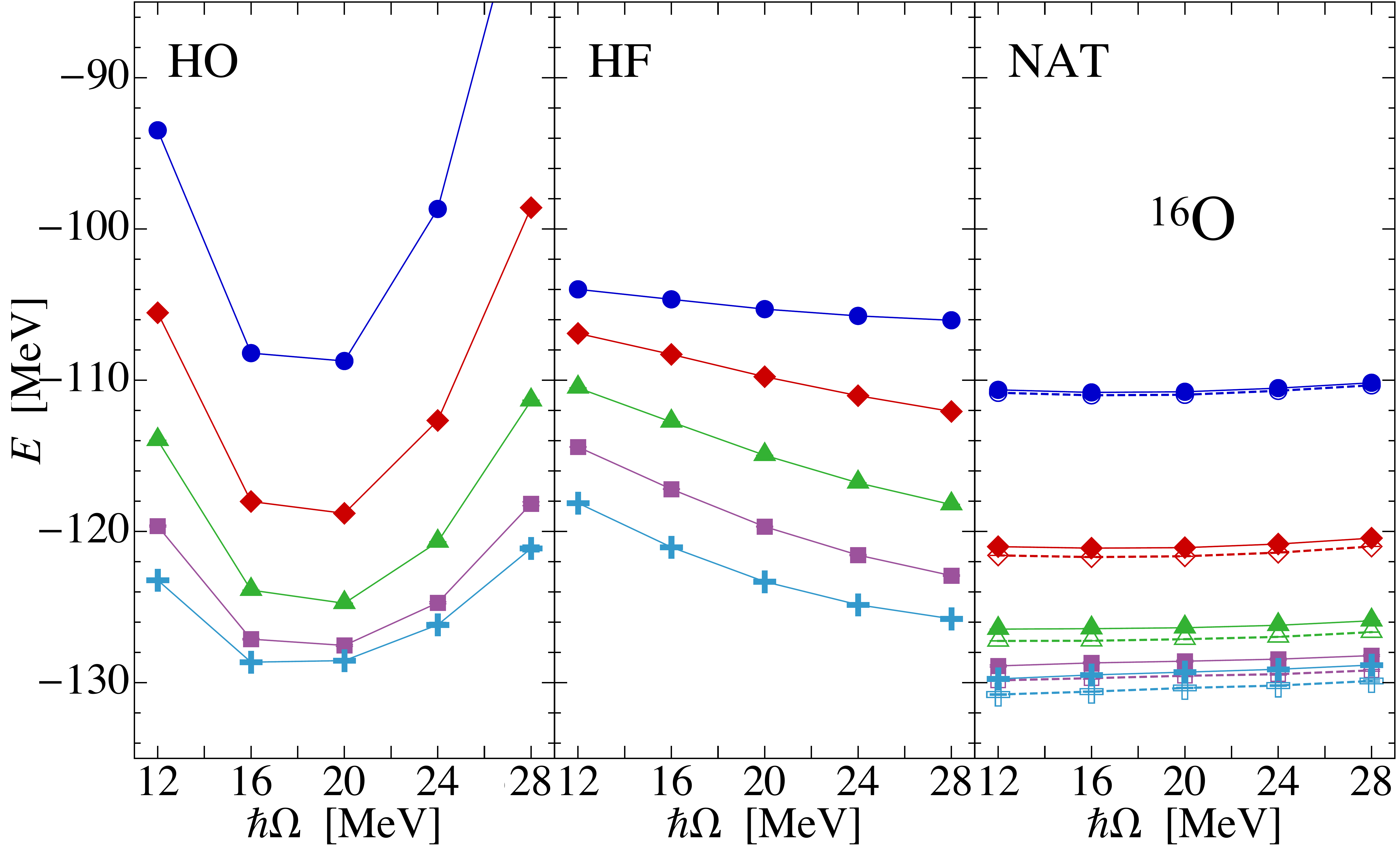}
\caption{(color online) Ground-state energies for \elem{O}{16} obtained in the NCSM with the HO, HF, and NAT basis sets with $N_{\max}=2$ (\symbolcircle[FGBlue]), $4$ (\symboldiamond[FGRed]), $6$ (\symboltriangle[FGGreen]), $8$ (\symbolbox[FGViolet]), and $10$ (\symbolcross[FGLightBlue]). The open symbols indicated results obtained with the NO2B approximation. All other paramters as in Fig.~\ref{fig:He4egs}.}
\label{fig:O16egs}
\end{figure}

Based on the differences of the wavefunctions among the HO, HF and NAT basis sets, we should expect quite different convergence patterns in actual NCSM calculations. We start with the discussion of the ground-state energy as function of the oscillator frequency $\hbar\Omega$ for increasing the model-space truncation parameter $N_{\max}$. 
The ground-state energies obtained for the three basis sets with the aforementioned chiral NN+3N interactions for \elem{He}{4} and \elem{O}{16} are summarized in Figs. \ref{fig:He4egs} and \ref{fig:O16egs}, respectively.
The left-hand panels depict the results with the HO basis, which exhibit a well-known systematics: At small $N_{\max}$ the energies show a pronounced minimum as function of frequency, often used to define an optimal HO frequency, with a monotonic rise of the energy towards smaller and larger frequencies. With increasing $N_{\max}$ the frequency-dependence gradually flattens and the minimum broadens. Once convergence is reached, the energies become independent of frequency and successive $N_{\max}$ calculations fall on top of each other. For \elem{He}{4} we reach this convergence already at $N_{\max}=10$, while for \elem{O}{16} still larger $N_{\max}$ would be needed. 

For the HF basis, shown in the middle panel, the pattern is vastly different. Obviously, we cannot extract meaningful converged results from the HF-basis calculation. There is a strong frequency dependence over the full $N_{\max}$-range considered here, favouring large frequencies. Furthermore, there is no indication of convergence even for model-space sizes where the HO calculation is converged already. This catastrophic behaviour results from a combination of two effects: the pathologies of the HF single-particle basis, as discussed above, and the $N_{\max}$ model space truncation. In a full CI calculation, which solely uses a truncation of the single-particle basis, the HF-basis would not pose a problem, because the full CI model space covers all possible unitary transformations of the single-particle states. Thus, the deficiencies of the HF-basis can be remedied by the full CI solution. This is not possible in a NCSM calculation, since we use a truncation in a many-body energy parameter and unitary transformations of the single-particle basis would lead beyond the $N_{\max}$-truncated space. 
Methods like coupled-cluster theory or the in-medium similarity renormalizaion group, which employ a truncation of the single-particle basis (i.e. an $e_{\max}$ truncation) plus some cluster truncation, can work with the HF basis, because they have the freedom to improve the single-particle basis (singles amplitudes in coupled cluster). However, this is only true as long as the HF calculation and the subsequent full CI or coupled-cluster calculations use the same single-particle truncation. 

For the NAT basis, the situation improves drastically. Already for small $N_{\max}$ there is almost no frequency dependence, which is expected based on the robustness of single-particle wavefunctions. With increasing $N_{\max}$ we observe  rapid and very smooth convergence---the convergence rate for all frequencies is as good as or even better than for the HO basis at its optimal frequency. These two aspects, frequency independence and optimal convergence, make the NAT basis a perfect tool for efficient NCSM calculations. We only have to consider a single $N_{\max}$-sequence for one standard frequency, instead of multiple sequences to determine the optimal frequency.

As mentioned earlier, the NAT basis does not guarantee a formal separation of intrinsic and center-of-mass motion and we have to expect center-of-mass contaminations. We monitor the expectation value of the HO center-of-mass Hamiltonian in all calculations. In all cases presented here, this expectation values is on the order of $100\,\text{keV}$ and it systematically decreases with $N_{\max}$. Therefore, we consider the effects of center-of-mass contaminations negligible. 

So far, we have discussed calculations including explicit 3N interactions in the Hamiltonian. We can simplify these calculations significantly using the normal-ordered two-body approximation (NO2B) for the 3N interaction, see Ref. \cite{RoBi12} for details. 
The normal-ordering of the 3N interaction with respect to a single-determinant reference state can be conveniently combined with the transformation of the interaction matrix elements into the NAT basis. This reduces the computational cost for the basis transformation process and the subsequent NCSM calculation significantly. Results for the \elem{O}{16} ground-state energy obtained with the NAT basis and the NO2B approximation for the 3N terms are indicated by the open symbols in the right-hand panel of Fig.~\ref{fig:O16egs}. We observe, in agreement with our earlier work \cite{RoBi12,BiLa13}, that the NO2B approximation produces an overbinding of about 1 MeV or 1\% compared to the calculations with explicit 3N interactions in large $N_{\max}$, which is an acceptable uncertainty for many applications beyond the lightest nuclei.

\subsection{Charge Radii} 

\begin{figure}[t!]
\centering
\includegraphics[width=0.95\columnwidth]{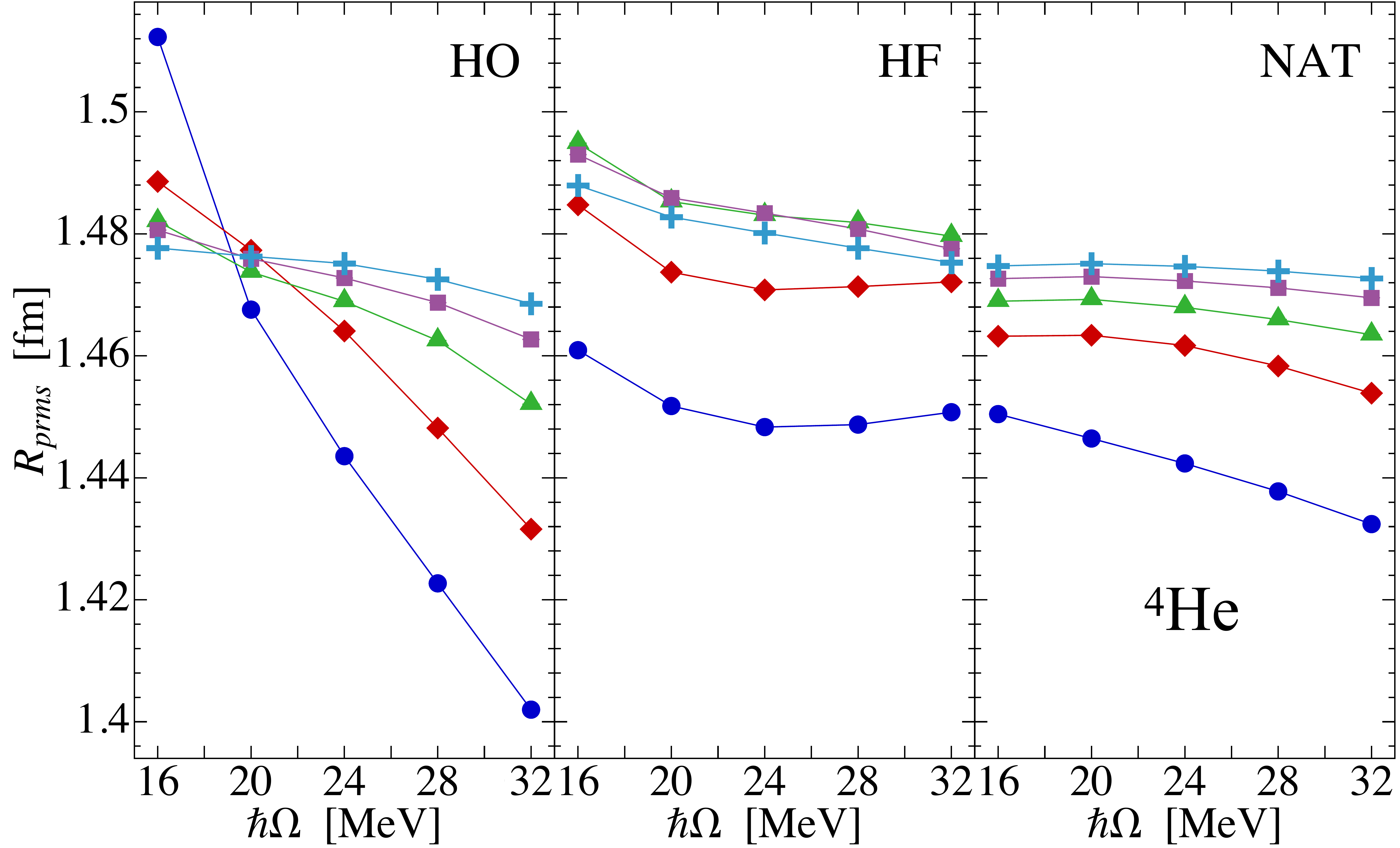}
\caption{(color online) Point-proton radii for the ground state of \elem{He}{4} obtained in the NCSM with the HO, HF, and NAT basis sets (panels from left to right) as function of the oscillator frequency $\hbar\Omega$ for $N_{\max}=4$ (\symbolcircle[FGBlue]), $6$ (\symboldiamond[FGRed]), $8$ (\symboltriangle[FGGreen]), $10$ (\symbolbox[FGViolet]), and $12$ (\symbolcross[FGLightBlue]). All calculations employ the chiral NN+3N interaction ($\Lambda_{\text{3N}}=400\,\text{MeV}/c$) after an SRG evolution with $\alpha = 0.08\,\text{fm}^4$.}
\label{fig:He4rad}
\end{figure}
\begin{figure}[t!]
\centering
\includegraphics[width=1\columnwidth]{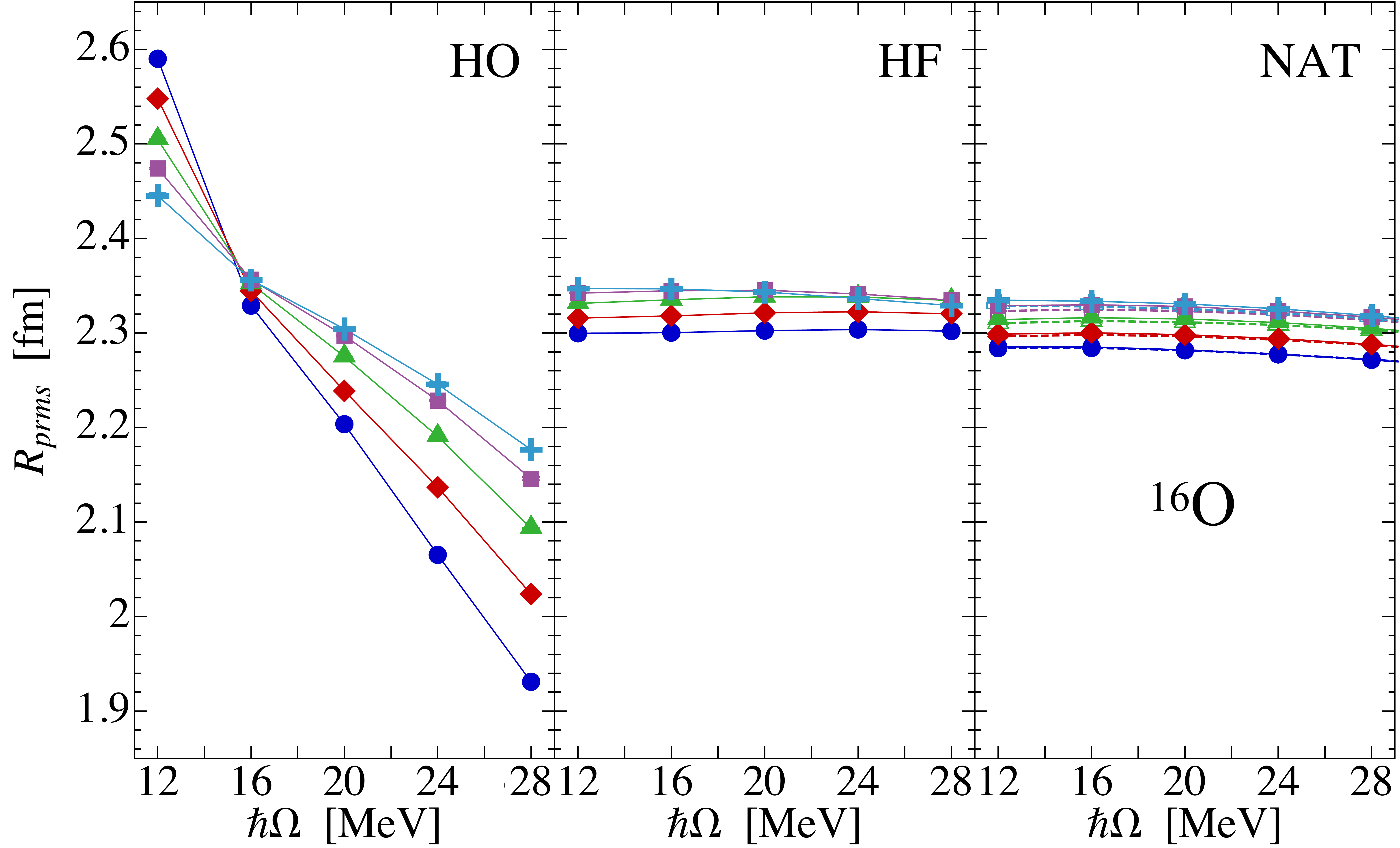}
\caption{(color online) Point-proton radii for the ground state of \elem{O}{16} obtained in the NCSM with the HO, HF, and NAT basis sets with $N_{\max}=2$ (\symbolcircle[FGBlue]), $4$ (\symboldiamond[FGRed]), $6$ (\symboltriangle[FGGreen]), $8$ (\symbolbox[FGViolet]), and $10$ (\symbolcross[FGLightBlue]). All other paramters as in Fig.~\ref{fig:He4rad}.}
\label{fig:O16rad}
\end{figure}

From the convergence point of view, root-mean-square radii are notoriously difficult observables. They are sensitive to the long-range behavior of the basis functions and should benefit from basis optimizations beyond the HO. Moreover, they are not protected by the variational principle and can exhibit a complicated, non-monotonous convergence pattern.

In analogy to the discussion of ground-state energies, Figs.~\ref{fig:He4rad} and \ref{fig:O16rad} show the point-proton radii for the ground states of \elem{He}{4} and \elem{O}{16}, respectively, as function of $\hbar\Omega$ for a sequence of $N_{\max}$ truncation parameters. For the HO basis, in the left-hand panels of both figures, we observe a characteristic pattern: At small $\hbar\Omega$---corresponding to large oscillator lengths---the radius decreases with increasing $N_{\max}$ while at large $\hbar\Omega$---corresponding to small oscillator lengths---the radius increases with $N_{\max}$. This leads to a flattening of the frequency-dependence, although full convergence is not reached in the present examples. To extract an estimate for the converged radius, many NCSM applications use the frequency where all $N_{\max}$ curves seem to intersect, i.e., an optimal frequency where the $N_{\max}$ dependence is minimized \cite{BoFu08}.   

For the NAT basis, depicted in the right-hand panels of Figs.~\ref{fig:He4rad} and \ref{fig:O16rad}, we again find a significant change in the convergence pattern: the frequency dependence is practically eliminated and we observe a very regular and monotonic convergence from below for all frequencies. Interestingly, the HF basis leads to a similar convergence pattern despite the pathological convergence behavior of the energy. The over-all change in the point-proton radius from the smallest to the largest model space is quite small, so that a robust and accurate extraction of radii is feasible. As for the energies, the main advantage of the NAT basis over the HO is the elimination of the oscillator frequency as a relevant parameter in the calculation. 

\subsection{Electromagnetic Observables}

\begin{figure}[t!]
\includegraphics[width=1.0\columnwidth]{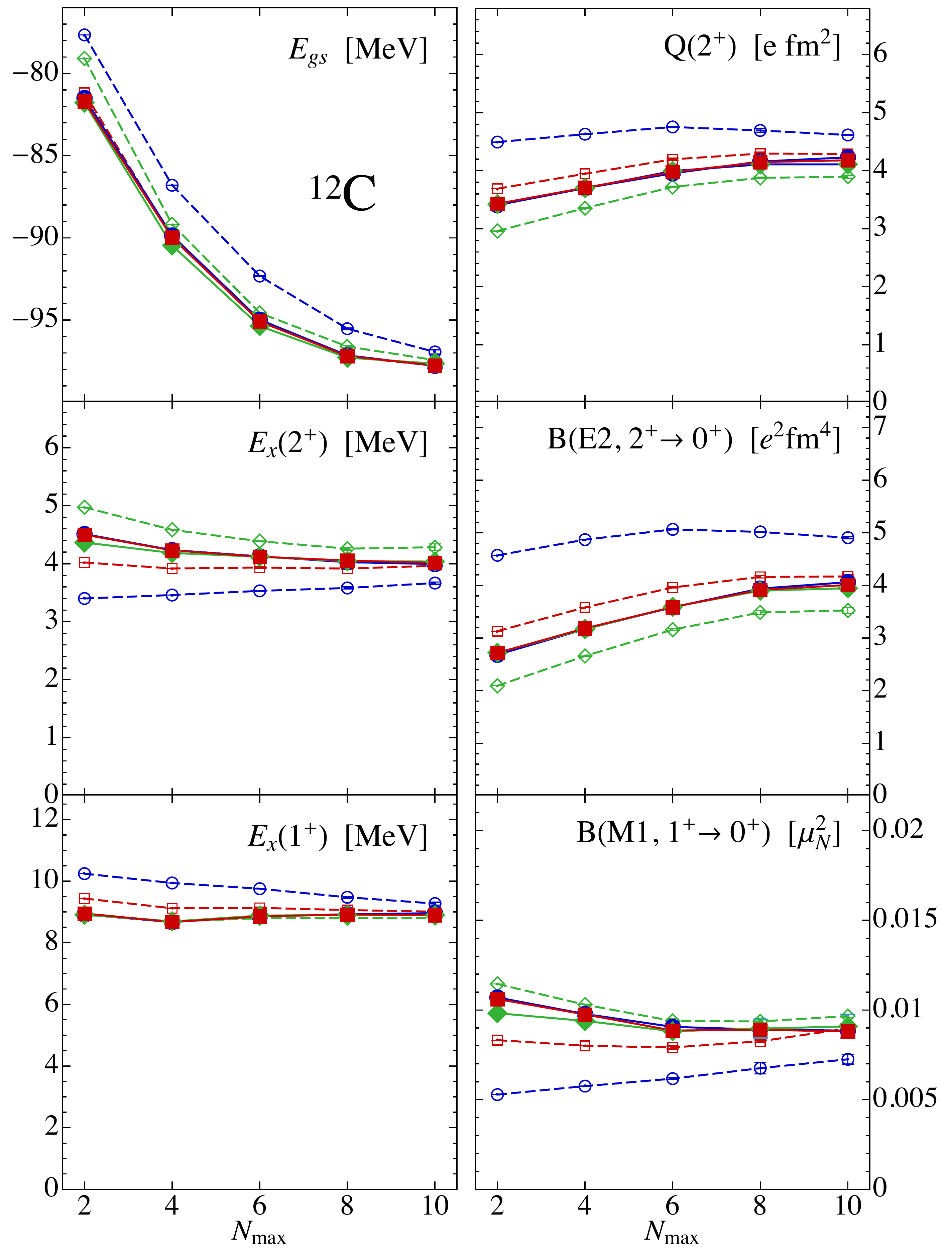}
\caption{(color online) Summary of the spectroscopy of \elem{C}{12} obtained in NCSM calculation with the HO basis (open symbols) and NAT basis (solid symbols) as function of $N_{\max}$ for $\hbar\Omega=16,\text{MeV}$ (blue), $20\,\text{MeV}$ (red), and $24\,\text{MeV}$ (green). The left-hand panels show the ground-state energy and the excitation energies for the first $2^+$ and $1^+$ states. The right-hand panels show the quadrupole moments of the first $2^+$ state, the $B(E2)$ transition strength from the $2^+$ to the ground state, and the $B(M1)$ strength from the $1^+$ to the ground state. All calculations employ the chiral NN+3N interaction ($\Lambda_{\text{3N}}=500\,\text{MeV}/c$) after an SRG evolution with $\alpha = 0.08\,\text{fm}^4$.}
\label{fig:EMobs}
\end{figure}

As a final check, we discuss excited states and electromagnetic observables, which are a prime target for applications of the NCSM. We consider \elem{C}{12} with the same chiral NN+3N interaction that was used for a detailed benchmark of the NCSM with and without importance truncation for a large range of spectroscopic observables \cite{MaVa14}. Figure~\ref{fig:EMobs} summarizes the results obtained with the HO and NAT basis for a suite of observables, including the ground-state energy, the excitation energy of the first $2^+$ and $1^+$ states, the quadrupole momentum of the first $2^+$ state, the B(E2) transition strength from the $2^+$ to the ground state, and the B(M1) transition strength from the $1^+$ to the ground state. All calculations were performed for three different oscillator frequencies and are presented as function of $N_{\max}$.

As in the previous cases, the NAT basis shows practically no frequency dependence for all observables and all model-space sizes. Thus, the main advantage of the NAT basis holds for excited states and spectroscopic observables as well---in practical calculations we only have to consider a single $\hbar\Omega$ parameter. Similar to the radii, the $N_{\max}$ dependence is very weak and generally monotonic. The largest model spaces used here, show clear indications of convergence for all the observables.
With the HO basis, there is a nontrivial dependence on the oscillator frequency in all cases. By tuning the oscillator frequency one can typically find a HO basis with a convergence behavior similar to the NAT basis. For all observables, the NAT results are fully consistent with the trends emerging from the HO calculations.

\section{Oxygen Isotopic Chain}

\begin{figure}[t!]
\centering
\includegraphics[width=1.0\columnwidth]{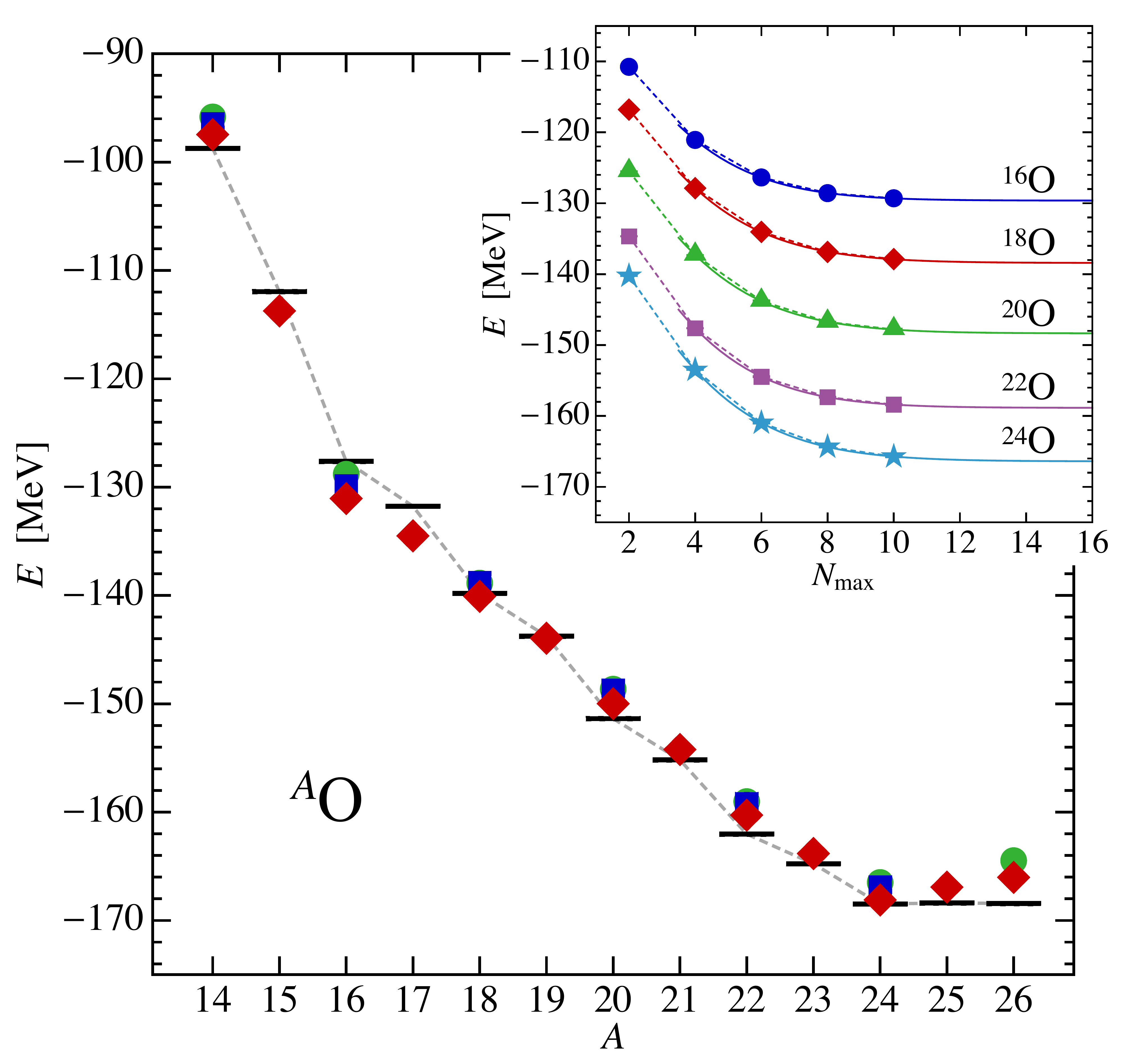}
\caption{(color online) Ground-state energies of all oxygen isotopes from \elem{O}{14} to \elem{O}{26} using the chiral NN+3N interaction ($\Lambda_{\text{3N}}=400\,\text{MeV}/c$) after an SRG evolution with $\alpha = 0.08\,\text{fm}^4$. Shown are NCSM calculations using the NAT basis ($\hbar\Omega=20\,\text{MeV}$) with the explicit 3N interaction (\symbolbox[FGBlue]) and with the NO2B approximation (\symboldiamond[FGRed]). For comparison we also plot the HO basis calculations of Ref.~\cite{HeBi13} with optimized oscillator frequency (\symbolcircle[FGGreen]). The inset shows the $N_{\max}$ dependence and extrapolation for selected isotopes with the explicit 3N interaction (see text).}
\label{fig:Oupdate}
\end{figure}

We conclude our numerical investigations, with an application of the NAT basis in large-scale NCSM calculations of the ground-state energies throughout the oxygen isotopic chain, including all even and odd isotopes. We have presented a first set of NCSM calculations for the even oxygen isotopes in Ref. \cite{HeBi13} using the same chiral NN+3N interaction. Those calculations were performed in the HO basis up to $N_{\max}=12$ using a careful optimization of $\hbar\Omega$ to obtain optimal convergence for each isotope. With the NAT basis, we use a standard frequency $\hbar\Omega=20\,\text{MeV}$ and model spaces only up to $N_{\max}=10$.

In Fig.~\ref{fig:Oupdate} we summarize the ground-state energies for all isotopes from \elem{O}{14} to \elem{O}{26} using explicit 3N interactions as well as the NO2B approximation. For comparison, our old results from Ref. \cite{HeBi13} are also included in the plot. The inset in Fig.~\ref{fig:Oupdate} shows the $N_{\max}$ dependence of selected isotopes to illustrate the excellent convergence. Technically, we still use an exponential extrapolation in $N_{\max}$ to extract the converged ground-state energy shown in the main plot, however, the difference between the $N_{\max}=10$ results and the extrapolated value, serving as a conservative proxy for the extrapolation error, is smaller than the symbol size. The NO2B approximation leads to a systematic lowering of the ground-state energy by 1\%, irrespective of the closed- or open-shell nature of the nucleus. The NAT calculations confirm and improve our original HO-basis NCSM results---at a much reduced computational cost since we only consider one frequency and model spaces up to $N_{\max}=10$.

\section{Summary and Outlook}

We have introduced natural orbitals extracted from a correlated one-body density matrix obtained in second-order MBPT as a very convenient and efficient single-particle basis for NCSM calculations. Our approach provides NAT basis sets optimized for the nucleus and Hamiltonian of interest at very low computational cost. NCSM calculations using this NAT basis show superior model-space convergence and independence of the frequency of the underlying oscillator basis for all relevant observables. Thus, in contrast to traditional NCSM calculations in the HO basis, we only have to consider one frequency and we can typically stop at smaller $N_{\max}$. We demonstrated these benefits for the oxygen isotopic chain from \elem{O}{14} to \elem{O}{26}. In addition, we highlighted some pathologies of the HF basis, which inhibit meaningful calculations in combination with $N_{\max}$-truncated model spaces. 

The application of MBPT-based natural orbitals is not restricted to the NCSM domain. The NAT basis and the transformed matrix elements can be constructed also for medium-mass and heavy nuclei. Therefore, medium-mass methods that presently use the HF basis can readily employ the NAT basis as well. This is particularly relevant for hybrid approaches, like the in-medium NCSM \cite{GeVo17} and the perturbatively improved NCSM \cite{Ti18}, which rely on the $N_{\max}$ truncation in parts of the calculation.

\section*{Acknowledgments}

This work is supported by the framework of the Espace de Structure et de r\'eactions Nucl\'eaires Th\'eorique (ESNT) at CEA, the  Deutsche Forschungsgemeinschaft (DFG, German Research Foundation) Projektnummer 279384907, SFB 1245, the Helmholtz International Center for FAIR (HIC for FAIR), and the BMBF through contracts 05P15RDFN1 and 05P18RDFN1 (NuSTAR.DA). Numerical calculations have been performed at the computing center of the TU Darmstadt (lichtenberg).

\end{document}